\documentclass[mathleft,fleqn,%
]{an1}
\def \e    {$\pm$} 
\def \es   {$\pm \; \sigma_{rms}$} 

\usepackage{graphicx}
\usepackage[varg]{txfonts}
\begin{document}

\Pagespan{1}{}
\Yearpublication{...}%
\Yearsubmission{2017}%
\Month{0}%
\Volume{...}%
\Issue{0}%
\DOI{xxxx.201700000}%

\title{Discovery of optical flickering from the symbiotic star EF Aquilae
}

\author{R.\,K.\,Zamanov\inst{1}\fnmsep\thanks{Corresponding authors:
        {rkz@astro.bas.bg,  kstoyanov@astro.bas.bg}}
\and  S.\,Boeva \inst{1}
\and  Y.\,M.\,Nikolov\inst{1}
\and  B.\,Petrov\inst{1}
\and  R.\,Bachev\inst{1}
\and  G.\,Y.\,Latev\inst{1}
\and  V.\,A.\,Popov\inst{2}
\and  K.\,A.\,Stoyanov\inst{1}
\and  M.\,F.\,Bode\inst{3}
\and  J. Mart\'i\inst{4}
\and  T. Tomov\inst{5}  
\and  A. Antonova\inst{6}
}
\titlerunning{Flickering from the symbiotic star EF Aql}
\authorrunning{Zamanov, Boeva, Nikolov  et al.}
\institute{
Institute of Astronomy and National Astronomical Observatory, Bulgarian Academy of Sciences,  Tsarigradsko Shose 72, 
1784  Sofia, Bulgaria 
\and 
IRIDA, National Astronomical Observatory Rozhen, 4700 Smolyan, Bulgaria
\and 
Astrophysics Research Institute, Liverpool John Moores University, IC2 Liverpool Science Park, Liverpool, L3 5RF, UK
\and 
Departamento de F\'isica (EPSJ), Universidad de Ja\'en, Campus Las Lagunillas, A3-420, 23071, Ja\'en, Spain 
\and 
Centre for Astronomy, Faculty of Physics, Astronomy and Informatics, Nicolaus Copernicus University, Grudziadzka 5, PL-87-100 Torun, Poland
\and 
Department of Astronomy, Faculty of Physics, St Kliment Ohridski University of Sofia, 5 James Bourchier Boulevard, 1164 Sofia, Bulgaria 
}

\received{2017 February 16}
\accepted{...}
\publonline{...}


\abstract{%
We report optical CCD photometry of the recently identified symbiotic star 
EF Aql. Our observations in Johnson V and B bands clearly 
show the presence of stochastic light variations with an amplitude of about 0.2 mag on a time scale of minutes.  
The observations point toward a white dwarf as the hot component in the system.   
It is the 11-th object among more than 200 symbiotic stars known with detected optical flickering. Estimates of the mass accretion rate onto the WD and the mass loss rate in the wind of the Mira secondary star lead to the conclusion that less than 1\% of the wind is captured by the WD. Eight further candidates for the detection of flickering in similar systems are suggested.
}

\maketitle

\section{Introduction}
EF Aquilae was identified as a variable star on photographic plates from 
K\"onigstuhl Observatory almost a century ago (Reinmuth 1925). 
It is associated with a bright 
infrared source -- IRAS 19491-0556 / 2MASS J19515172-0548166.  
Le Bertre et al. (2003) provide K and L' photometry for EF Aql, and classify it as an oxygen-rich
asymptotic giant branch star located at a distance of $d=3.5$~kpc and 
losing mass at a rate $3.8 \times 10^{-7}$ M$_\odot \: yr^{-1}$.
Richwine et al. (2005) have examined the optical 
survey data for EF~Aql and classify it as a Mira type variable 
with  a period of 329.4 d and amplitude of variability $> 2.4$ mag.  
The optical spectrum shows  prominent Balmer emission lines visible through at least H11
and $[O \: III]\: \lambda 5007$ emission. 
The emission lines and the bright UV flux detected in GALEX satellite images
provide undoubted evidence for the presence of a hot companion.
Thus EF Aql appears to be a symbiotic star,
a member of the symbiotic Mira subgroup (Margon et al. 2016). 

The symbiotic stars are  long-period interacting binaries, consisting of an evolved giant 
transferring mass to a hot compact object. 
Their orbital periods are in the range from 100 days to more than 100 years. 
A cool giant or supergiant of spectral class G-K-M is the mass donor.  
If this giant has Mira-type variability, the system usually is a strong infrared source. 
The hot secondary accretes material supplied from the red giant. 
In most symbiotic stars, the secondary is a degenerate star, typically a white dwarf or subdwarf. 
In a few cases has the secondary been shown to be 
a neutron star (e.g. Bahramian et al. 2014; Kuranov \& Postnov 2015;
and references therein).

Systematic searches  for  flickering variability in  symbiotic stars 
and related objects (Dobrzycka et al. 1996; 
Sokoloski, Bildsten \& Ho \ 2001;  Gromadzki et al. 2006; 
Stoyanov 2012; Angeloni et al. 2012, 2013)  have shown that 
optical flickering activity is rarely detectable.
Among more than 200 symbiotic stars known, 
only 10 present flickering -- RS Oph, T CrB, MWC 560, Z And, 
V2116 Oph, CH Cyg, RT Cru, {\it o} Cet, V407 Cyg, and V648 Car. 


Here we report optical CCD photometry of EF~Aql and detection of flickering
in Johnson V and B bands.

 \begin{figure*}   
 \vspace{7.0cm}   
  \includegraphics{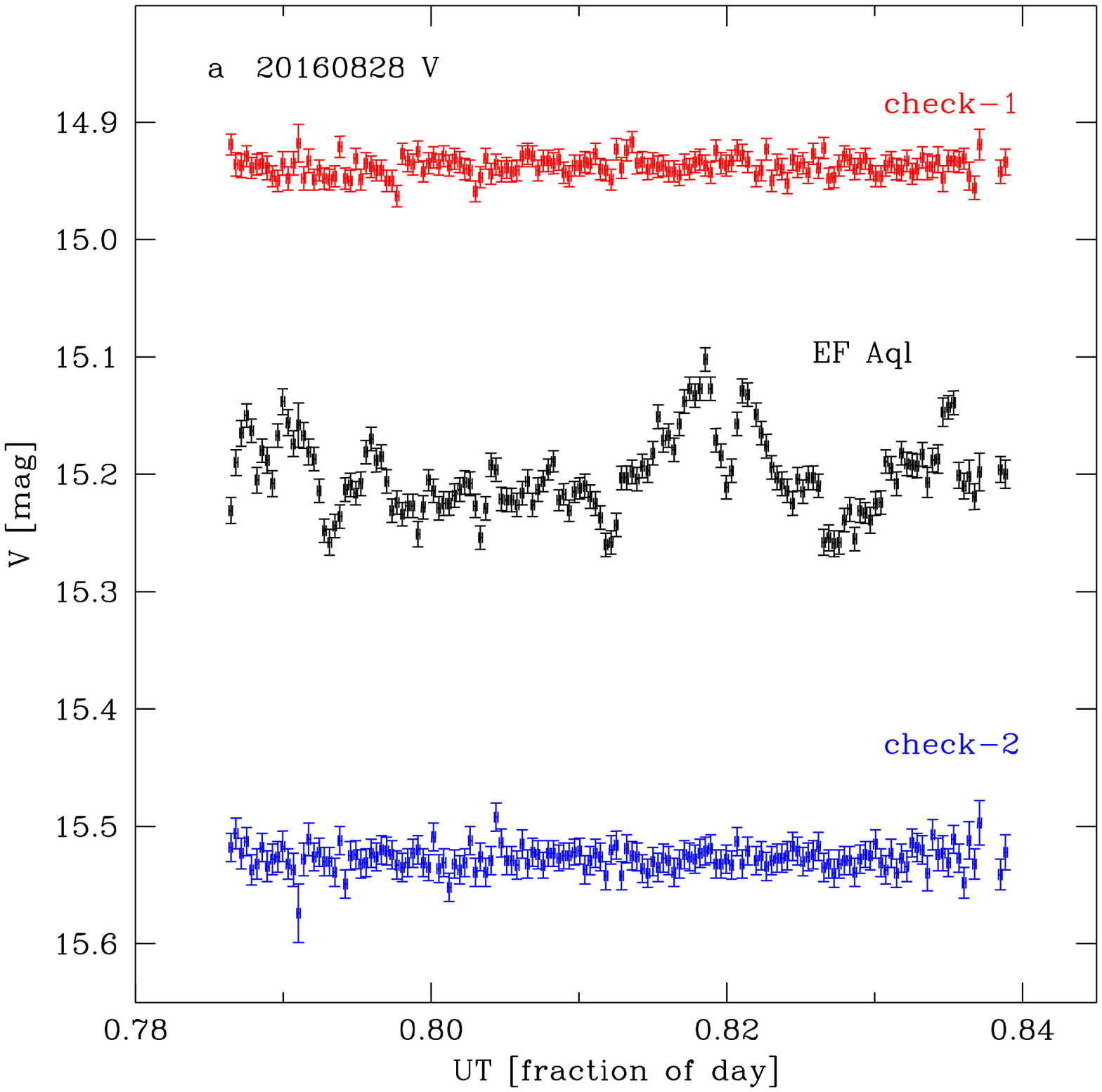}      
  \includegraphics{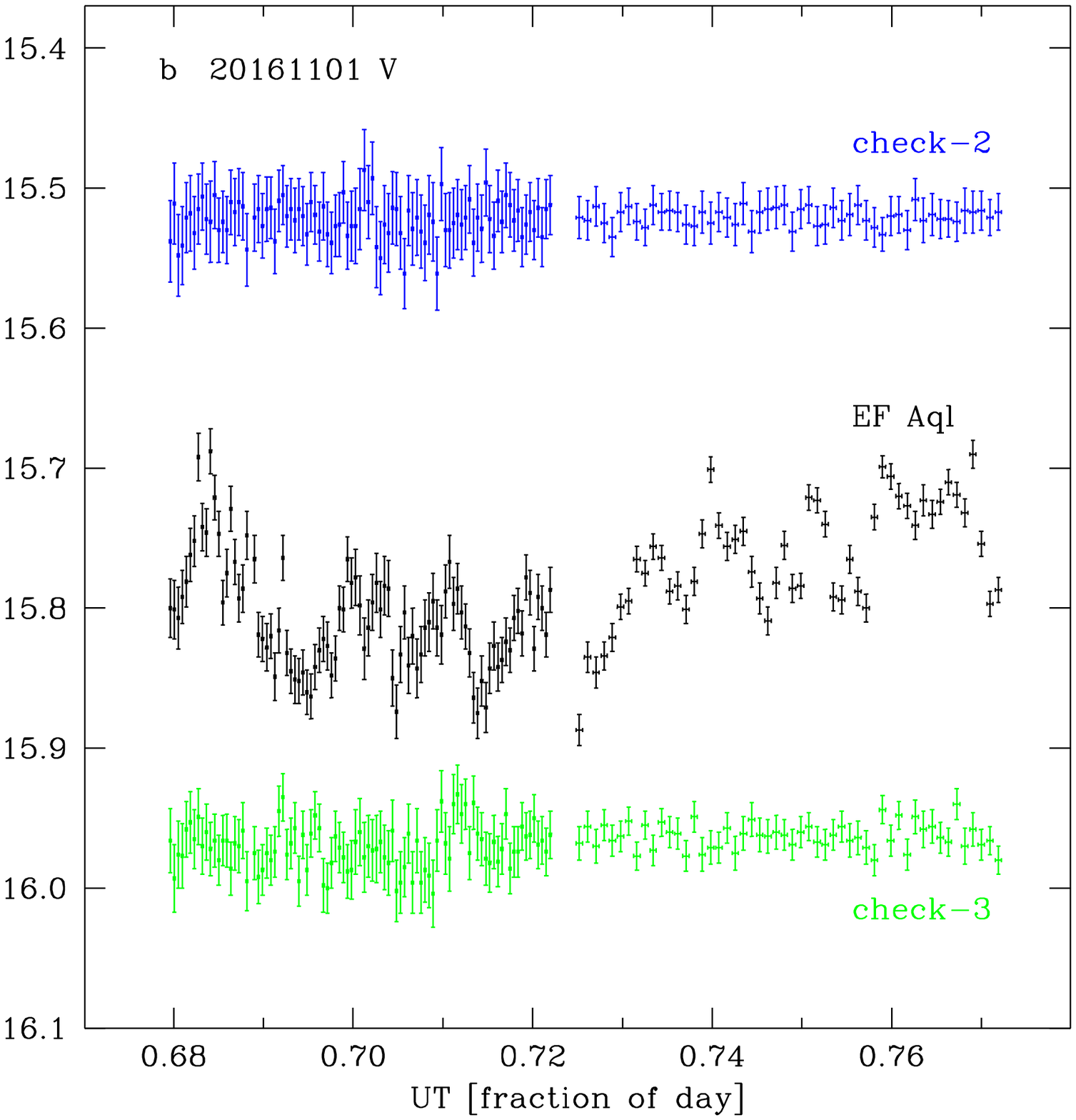}      
  \includegraphics{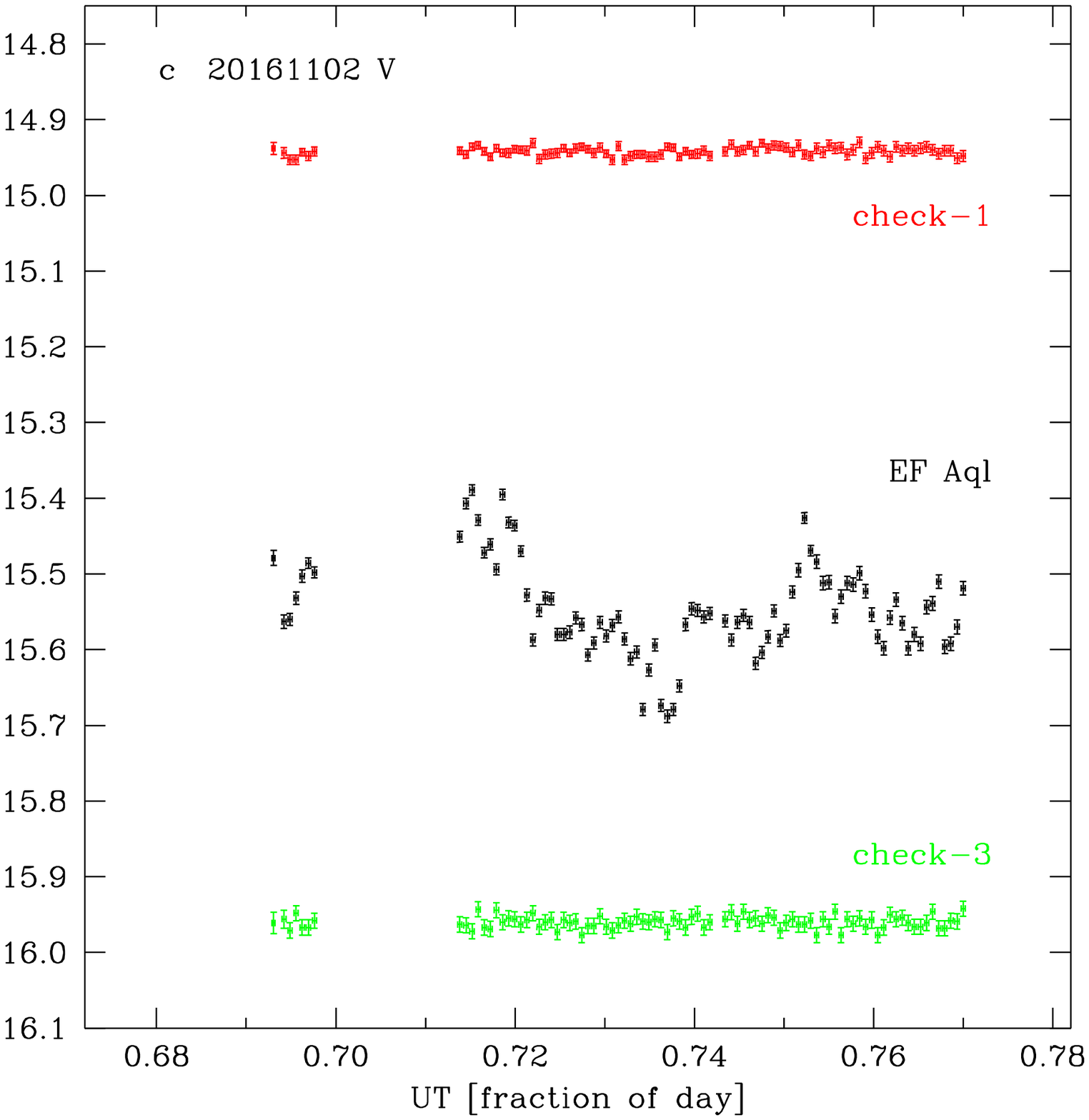}      
\caption[]{ Detection of optical flickering of EF Aql. In each panel two check stars 
are also shown on the same scale. 
It is clearly seen that EF Aql varies with an amplitude larger than 0.15 mag. 
}  
\label{fig.11}      
\end{figure*}	     

\begin{table*}
  \centering
  \caption{CCD observations of EF~Aql. In the table are given as follows:
the date of observation (in format yyyymmdd),  UT-start and UT-end of the run,
the telescope, band, exposure time, number of  CCD images obtained, 
minimum -- maximum magnitudes in each band,
average magnitude in the corresponding band, 
typical observational error.
}
  \begin{tabular}{lcc ccr ccc  cccccr} 
\hline
date &    UT       & telescope & filter & exp-time  & $N_{pts}$ & min-max   & average &   err  & \\
     & start - end &           &        &    [s]    &           & mag - mag & mag     &   mag  & \\
\hline
20160828 & 18:52 - 20:11  & 2.0 m       & V & 20s      & 146  & 15.102 - 15.260 & 15.201 &   0.010 & \\
20160828 & 18:53 - 20:08  & 50/70 cm    & B & 120s     &  32  & 16.053 - 16.233 & 16.144 &   0.020 & \\
20160903 & 19:42 - 20:55  & 41 cm Jaen  & I & 60s      &  36  & 10.920 - 10.944 & 10.932 &   0.002 & \\
20161028 & 16:27 - 17:54  & 60 cm Bel   & V & 120s     &  40  & 15.564 - 16.028 & 15.773 &   0.033 & \\
20161030 & 16:06 - 16:29  & 2.0 m       & V & 20s      &  32  & 15.371 - 15.473 & 15.412 &   0.009 & \\
20161101 & 16:18 - 18:32  & 2.0 m       & V & 20s, 60s & 150  & 15.690 - 15.809 & 15.752 &   0.010 & \\
20161102 & 16:36 - 18:29  & 2.0 m       & V & 40s      &  90  & 15.389 - 15.688 & 15.546 &   0.008 & \\
20161110 & 16:49 - 18:20  & 30 cm Irida & V & 150 s    &  33  & 15.494 - 15.888 & 15.705 &   0.046 & \\ 
\hline
  \end{tabular}
  \label{tab.obs}
\end{table*}

\begin{table*}
  \centering
  \caption{V band observations of  EF Aql and the check stars in the field. 
  The average magnitude and  $\sigma_{rms}$ are given for each object.  }
  \begin{tabular}{lcc ccc ccc | ccr} 
\hline
date  & telescope & filter & EF Aql          &    check-1       &  check-2         &   check-3          & \\
      &           &        & mean \es        &   mean \es       & mean \es         &  mean \es          & \\
\hline
20160828 & 2.0 m     & V & 15.201 \e\ 0.033  & 14.937 \e\ 0.008 & 15.527 \e\ 0.010 & 15.964 \e\ 0.013	& \\
20161028 & 60 cm Bel & V & 15.773 \e\ 0.104  & 14.986 \e\ 0.038 & 15.574 \e\ 0.023 & 16.062 \e\ 0.040	& \\
20161030 & 2.0 m     & V & 15.412 \e\ 0.030  & 14.914 \e\ 0.006 & 15.465 \e\ 0.008 & 15.948 \e\ 0.015	& \\
20161101 & 2.0 m     & V & 15.752 \e\ 0.035  & 14.937 \e\ 0.008 & 15.520 \e\ 0.007 & 15.962 \e\ 0.009	& \\  
20161102 & 2.0 m     & V & 15.546 \e\ 0.061  & 14.943 \e\ 0.007 & 15.525 \e\ 0.009 & 15.960 \e\ 0.009	& \\
20161110 & 30 cm     & V & 15.705 \e\ 0.094  & 14.936 \e\ 0.033 & 15.511 \e\ 0.034 & 15.976 \e\ 0.060   & \\
\hline  
  \end{tabular}
  \label{tab.sig}
\end{table*}

\section{Observations}

During the period August - November 2016, we secured  CCD photometric monitoring  
with  5 telescopes equipped with CCD cameras:
\begin{itemize}
\item the 2.0 m RCC telescope of the National Astronomical Observatory Rozhen, Bulgaria
(CCD VersArray 1300 B, $1340  \times  1300$ px);

\item the 50/70 cm Schmidt telescope of NAO Rozhen (SBIG STL11000M CCD, $4008 \times 2672$ px); 

\item the 60 cm telescope of the Belogradchick Astronomical Observatory 
(SBIG ST8 CCD, $1530 \times 1020$ px);  

\item  the 30 cm astrograph of IRIDA observatory (CCD camera ATIK 4000M, 
$2048\times2048$ px);   

\item the automated 41~cm                          
telescope of the University of Ja\'en, Spain - ST10-XME CCD camera with $2184 \times 1472$ px 
(Mart{\'{\i}}, Luque-Escamilla, \&  Garc\'{\i}a-Hern\'andez 2017).  
\end{itemize}
All the CCD images have been bias subtracted, flat fielded, and standard 
aperture photometry has been performed. The data reduction and aperture photometry 
are done with IRAF and have been checked with alternative software packages. 
A few objects from the APASS catalogue  (Munari et al. 2014; Henden et al. 2016) have been used
as comparison stars.  As check stars we used   USNO U0825.17150321,  USNO  U0825.17161750 and USNO  U0825.17157668. 
In Fig.~1 and Table~2, they are marked as check-1, check-2 and check-3, respectively.

The results of our observations are summarized in Table~\ref{tab.obs} and 
plotted on Fig.~\ref{fig.11}. 
For each run  we measure the minimum, maximum, and average brightness
in the corresponding band, plus the standard deviation of the run. 
For each run we calculate the root mean square (rms) deviation
\begin{equation}
    \sigma_{rms} = \sqrt{ \frac{1}{N_{pts}} \sum\limits_{i}(m_i - \overline m )^2 } ,
    \label{eq.sig}
\end{equation}
where $\overline m$ is the average magnitude in the run. 
$\sigma_{rms}$ calculated in this way includes the contributions 
from the variability of the star (if it exists) and from the measurement errors. 
For non-variable stars it is a measure of the accuracy of the photometry.

 \begin{figure}   
 \vspace{8.0cm}   
  \includegraphics{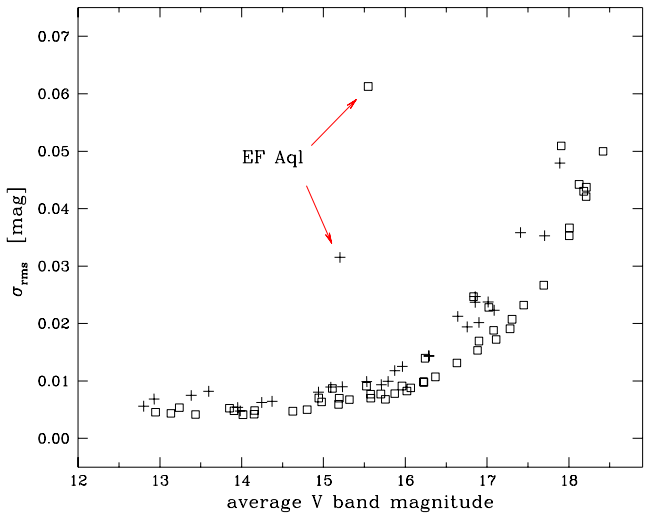}      
\caption[]{The root mean square deviation versus the average V band magnitude. 
The plus signs refer to the night 20160828, the squares -- to 20161102. 
In both nights, EF~Aql deviates considerably from the behaviour of the other stars, which indicates that 
it is variable during our observations. }  
\label{fig.2}      
\end{figure}	     

\section{Detection of flickering in EF Aql}

During our observations the V band brightness of EF~Aql was in the range $ 15.1 \le  V \le 16.03$. 
The General Catalogue of Variable Stars (Samus et al. 2017) suggests brightness of EF~Aql in the range
$12.4 \le V \le 15.5$. 
The ASAS data (Pojmanski \& Maciejewski 2004) show variability in the range $12.5 \le V \le 16.0$. 
The lowest brightness in our set is  $V \approx 16.03$, which is about 
the minimum brightness measured in ASAS, indicating that our observations are near the 
minimum of the Mira cycle.  

Rapid aperiodic brightness variations, like the flickering 
from cataclysmic variables  (Warner 1995, Bruch 2000), 
is evident on all our observations in B and V bands. 
It is not detectable in I band, suggesting that
the I flux is dominated by the red component. 
Fig.~\ref{fig.11} shows the light curves from three observations.  
For comparison, two check stars (one brighter than EF Aql, and one fainter)
are also plotted on the same scale. 
It is apparent that the variability of EF~Aql is considerably larger
than that of the check stars, which indicates that it is not a result of observational errors. 

The $\sigma_{rms}$ expected from the accuracy of the photometry can be deduced  
from the observations of the check stars with brightness similar to that of EF~Aql. 
In Table~2 we give the mean magnitude and   $\sigma_{rms}$ of EF~Aql and the three check stars. 
The root mean square deviation $\sigma_{rms}$  of EF~Aql
exceeded the rms deviation expected from the check star 
by more than a factor of 2.

In Fig.~\ref{fig.2} we plot $\sigma_{rms}$ for EF~Aql and of about 30 other stars from the field around EF~Aql. 
During run 20160828 (plotted with plus signs),  EF~Aql exhibits flickering with peak to peak amplitude 
0.16 mag. The root mean square deviation is $\sigma_{rms}(EF Aql) = 0.033$~mag. 
For stars with similar brightness 
we have $\sigma_{rms} \approx 0.009$~mag. In other words, $\sigma_{rms}$ of EF~Aql is more than three times 
larger than expected from observational errors. 
During run 20161102, EF~Aql exhibits flickering with higher amplitude of about 0.30 mag.
The root mean square variability is $\sigma_{rms}(EF Aql) = 0.061$~mag. For stars with similar brightness 
we have  $\sigma_{rms} \approx 0.008$~mag. In other words, the rms of EF~Aql is more than seven times 
larger than that expected from observational errors

Using  our simultaneous  B and V band observations obtained on 28 August 2016 (see Table~1) and 
interstellar extinction  $A_V=0.45$ (Margon et al. 2016),  
we calculate the dereddened colour of the flickering source as $(B-V)_0 = 0.35 \pm 0.05$. 
For comparison, the average $(B-V)_0$  colour of the flickering source 
is  $0.25 \pm 0.21$ in the recurrent novae T~CrB and RS~Oph, 
and  $0.10 \pm 0.20$ in the cataclysmic variables (Bruch 1992; Zamanov et al. 2015). 
It appears that  $(B-V)_0$ of the flickering source in EF~Aql is more similar to the 
flickering source in T~CrB and RS~Oph, which also contain a red giant mass donor.

\section{Discussion}

From our R and I band observations of EF~Aql, we measure $(R-I)=3.08 \pm 0.06$ and $(R-I) = 2.93 \pm 0.05$, for 
28 August 2016 and  4 September 2016, respectively. 
Applying  $A_V=0.45$ and the calibration of Fitzpatrick (1999), we correct $(R-I)$ for interstellar reddening,  
and find  $ 2.78 < (R-I)_0  < 3.04$, which (using the results of Celis 1984) corresponds to 
a spectral subtype of the asymptotic giant branch star of M7 - M8. 
An asymptotic giant of spectral type M7 - M8 is expected to have $(V-R)_0 \approx  4.0 - 4.5$ (Celis 1982)
and we estimated the brightness of the red giant in the V band around the time of our observations 
(at minimum of the Mira cycle) as $V \sim 17.8 - 18.5$ mag.  

The non-variable light from the red giant, 
contributes about 20\% of the flux  at V band. 
Fig.~1 demonstrates that the V band flux can change by more than 5\%  (0.05 mag) in less than 5 minutes
and more than 20\% (0.20 mag) in less than one hour. 
Taking into account the contribution of the red giant,
these rapid fluctuations correspond to variations up to 
$\pm 25$\%  (from the average level)
in the V-band  flux from the hot component of EF~Aql. 
The brightness fluctuations of EF~Aql are similar to those
observed in the prototype Mira (omicron Ceti) by Sokoloski \& Bildsten (2010). 

The flickering (stochastic photometric variations on timescales of
a few minutes with amplitude of a few$\times0.1$ magnitudes)
is a variability typical for the accreting white dwarfs in
cataclysmic variables and recurrent novae.  
About the nature of the hot companion in EF~Aql, Margon et al. (2016) supposed 
that the hot source is likely more luminous than a white dwarf, and thus may well be a subdwarf.
The persistent presence of minute-timescale stochastic optical
variations (see Table~1) with the observed amplitude is a strong indicator that
the hot component in EF~Aql is a white dwarf. 

A comparison of the flickering of EF~Aql with the flickering of the symbiotic recurrent nova RS~Oph 
(see Fig.~1 in Zamanov et al. 2010) shows that in RS Oph  
the flickering is visible in  BVRI bands, while 
in EF Aql it is not detectable in I, but well visible in  B and V bands. 
In  RS~Oph the mass accretion rate is of about 
$\sim 2 \times 10^{-8}$~M$_\odot$~yr$^{-1}$ (Nelson et al. 2011).
In EF Aql we see flickering in V, which means that the hot component is brighter than
the M giant in V band. Overall, the relative colour dependence of flickering in EF Aql implies that the mass accretion rate is lower than that in RS Oph, 
but not too much lower, probably of a few$\times 10^{-9}$~M$_\odot$~yr$^{-1}$.
Le Bertre et al. (2003) estimated that the mass donor in EF~Aql
is losing  mass  at a rate $3.8 \times 10^{-7}$ M$_\odot \: yr^{-1}$.
We used 2MASS K=5.36 mag and IRAS 12-micron flux (4.78 Jy) to estimate the 
color K-[12] defined by Gromadzki et al. (2009). 
Then, applying their Eq.~4, we determined a mass loss rate of $2\times 10^{-6}$~M$_\odot$~yr$^{-1}$ for EF~Aql.
This means that the white dwarf is capturing less than 1\% of the stellar wind of the red giant. 

In addition to the optical observations presented here, we searched the new gPhoton database (Million et al. 2016) 
for GALEX ultraviolet observations of EF Aql. gPhoton has a
calibration and extraction pipeline that allows easy access to calibrated
GALEX data. Using its module gFind, we found five epochs of observations
in the near UV band (NUV, 1771~\AA  -- 2831~\AA), only one of which was
previously reported by Margon et al. (2016). We then used
the gMap and gAperture modules to determine aperture size and background
annulus size and positions, to make sure all the counts are inside the
aperture and no contaminating sources are within the background
subtraction annulus. We thus obtain the following fluxes: \\
$ 20040630 \; 17:06 \;  \;  \;   6.25 \pm 0.09 \times 10^{-15}$ erg s$^{-1}$ cm$^{-2}$ \AA$^{-1}$,  \\
$ 20050627 \; 14:39 \;  \;  \;   6.64 \pm 0.08 \times 10^{-15}$ erg s$^{-1}$ cm$^{-2}$ \AA$^{-1}$,  \\
$ 20100813 \; 16:53 \;  \;  \;   7.98 \pm 0.04 \times 10^{-15}$ erg s$^{-1}$ cm$^{-2}$ \AA$^{-1}$,  \\
$ 20100815 \; 09:58 \;  \;  \;   8.26 \pm 0.04 \times 10^{-15}$ erg s$^{-1}$ cm$^{-2}$ \AA$^{-1}$,  \\
$ 20100815 \; 13:15 \;  \;  \;   8.37 \pm 0.04 \times 10^{-15}$ erg s$^{-1}$ cm$^{-2}$ \AA$^{-1}$,  \\
where the time  is given in the format  yyyymmdd~hh:mm. 
These results show that the NUV flux of EF Aql in August 2010 was $\approx
30\%$ larger than in June 2004, probably indicating variable mass
accretion rate onto the white dwarf.

There are more than 200 symbiotic stars  known (Belczy{\'n}ski et al. 2000).
Among them  flickering is detected in only 11 objects
(including EF~Aql), i.e. in 5\% of the cases (see Sect.~1 for references).  
On the basis of their infrared properties, 
the symbiotic stars are divided in three main groups S-, D-, and D'-type (Allen 1982;  Miko{\l}ajewska 2003). 
There are about 30 symbiotic stars classified as symbiotic Miras (Whitelock 2003). 
In three of them flickering is present, i.e.10\% of the objects. 
It seems that flickering more often can be detected in  symbiotic Miras 
than among S- and D'-type symbiotics. 
Bearing in mind this, we searched in the Catalogue of Symbiotic Stars (Belczy{\'n}ski et al. 2000),
for D-type symbiotics with low ionization potential which are potential candidates for flickering detection. 
In our opinion  V627~Cas, KM~Vel, BI~Cru, V704~Cen, Hen~2-139, V347~Nor, 
WRAY~16-312 and LMC~1 deserve to be searched for flickering 
near the minima of the Mira brightness cycle.

Symbiotic binaries (especially those with detected rapid variability) have historically revealed remarkable events of      
acceleration of ejection of collimated outflows (Taylor et al. 1986; Crocker et al. 2002; Brocksopp et al. 2004),
although not as powerful as in X-ray binaries and microquasars. Nevertheless,
some of their collimated ejecta display a combination of thermal and non-thermal
emission mechanisms (Eyres et al. 2009). 
Inspection of the EF~Aql position in the NRAO VLA Sky Survey (NVSS, Condon et al. 1998) reveals no radio source
detection with a $3\sigma$ upper limit of 1.6 mJy at 20 cm. As a comparison, this upper limit is very similar to the
faint source detection of the nearby symbiotic binary system CH Cygni that appears at a flux density level of 2.9 mJy in
the same survey. However, EF Aql is significantly more distant than CH Cygni (at least 15 times) and therefore the lack
of detection at the NVSS sensitivity is not surprising. Deeper radio mapping with modern interferometers would be thus
desirable. With the similarities between RS Oph and EF Aql noted above, it may be worthwhile 
to look for possible recurrent nova outbursts of the latter system in archival data, 
as has been done for other objects (e.g. Schaefer 2010).


\section{Conclusions}
On seven nights during the period August - November 2016, we performed 11.2 hours 
photometric 
observations of the symbiotic Mira EF~Aql.  
We find that EF~Aql exhibits short-term optical variability (flickering) on a time scale of minutes. 
The detected amplitude is about 0.15 - 0.30 mag  in B and V bands. 
The root mean square deviation of EF~Aql is from three to seven times 
larger than that expected from observational errors.   
The presence of flickering strongly suggests that the hot component is a white dwarf.
It seems that the flickering is more often seen in symbiotic Miras than in the general population 
of symbiotic stars.

\acknowledgements
This work was partly supported by grants  DN 08/1~13.12.2016 (Bulgarian National Science Fund),  
AYA2016-76012-C3-3-P (Spanish Ministerio de Econom\'{\i}a y Competitividad, MINECO)
as well as FEDER funds.

\end{document}